\documentclass{elsart}
\usepackage{epsfig}
\usepackage{amssymb}
\usepackage{amsmath}
\newcommand{\dla}{\langle{\!}\langle}
\newcommand{\dra}{\rangle{\!}\rangle}
\newcommand{\bigdla}{\left<{\!}\left<}
\newcommand{\bigdra}{\right>{\!}\right>}

\begin{document}

\begin{frontmatter}



\title{A $p$-Spin Interaction Ashkin-Teller Spin-Glass Model}
\author[1,2]{I.S. Queiroz Jr.}\ead{\\ \quad\quad idalmir@dfte.ufrn.br},
\author[1]{F.A. da Costa\corauthref{cor1}}\ead{\\ \quad\quad
fcosta@dfte.ufrn.br}, and
\author[1]{F.D. Nobre}\ead{\\ \quad\quad nobre@dfte.ufrn.br}
\address[1]{Departamento de F\'{\i}sica Te\'orica e Experimental,
Universidade Federal do Rio Grande do Norte, 59072-970, Natal, RN, Brazil}
\address[2]{Departamento de F\'{\i}sica,
Universidade do Estado do Rio Grande do Norte, 59600-970, Mossor\'o, RN, Brazil}
\corauth[cor1]{Corresponding author}

\begin{abstract}
A $p$-spin interaction Ashkin-Teller spin glass, with three independent
Gaussian probability distributions for the exchange interactions, is studied
by means of the replica method.
A simple phase diagram is obtained within the replica-symmetric
approximation, presenting an instability of the paramagnetic solution at
low temperatures. The replica-symmetry-breaking procedure is implemented
and a rich phase diagram is obtained; besides the paramagnetic phase, three
distinct spin-glass phases appear. Three first-order critical frontiers are
found and they all meet at a triple point; among such lines, two of them
present discontinuities in the order parameters, but no latent heat, whereas
the other one exhibits both discontinuities in the order parameters and a
finite latent heat.

\end{abstract}


\end{frontmatter}

\section{Introduction}
\label{intro}

In recent years much progress has been achieved in the understanding of
magnetic disordered systems. Among those, one may single out the spin glasses
(SGs) \cite{bin,fischerhertz,youngbook}, for which significant
advances were obtained, as a result of a large effort dedicated to them.
Most of the SG studies were carried for two-spin-interaction models, that
were at the early stage, intended to explain the physical behavior of some
peculiar magnetic compounds. Nowadays, apart from this motivation, such
SG models were identified to be closely related to a large diversity of
physical problems, like neural networks, protein folding, and optimization
problems \cite{fischerhertz,youngbook}. The mean-field theory of the Ising
SG model, considered in terms of an infinite-range-interaction model
\cite{sk}, is nowadays accepted as satisfactorily understood. The simplest
solution, based on a single SG order parameter, known as
replica-symmetric (RS) solution \cite{sk},
presented serious difficulties, and it was shown to be unstable at low
temperatures \cite{alt}. The correct mean-field solution for the Ising SG
was proposed by Parisi \cite{par}, and it consists in an infinite
number of order parameters -- a procedure called of replica-symmetry breaking
(RSB). However, it is not clear up to the moment, whether the mean-field
solution is appropriate for the description of real --
short-range-interaction -- SG systems \cite{youngbook}.

The $p$-spin-interaction ($p>2$) models were initially introduced as merely
theoretical problems \cite{der80,der81}; however, the identification of a
close analogy between the transitions occurring
in p-spin interaction SG models and those obtained in mode coupling
theories of structural glasses \cite{kt1,kt2}, motivated many studies
on such models. A nice feature of infinite-range p-spin-interaction SG
models is that in the $p\rightarrow \infty$ limit the energy levels become
independent random variables, yielding an exactly solvable model, known as
the random-energy model \cite{der80,der81}.
Also, for $p>2$ the phase transition scenario of these SG models is
quite different from the one found in the corresponding $p=2$
model. Besides the usual equilibrium transition temperature, there
exists another transition temperature presenting a dynamical character. Right
below the equilibrium transition, a single step in the RSB procedure is
sufficient for stabilility \cite{gm}, i.e., the order parameter is properly
represented in terms of a one-step Parisi function.
For models where there is some kind of
competition between different types of interactions, the one-step RSB order
parameter changes dramatically the phase diagrams obtained from RS
solutions \cite{mot,jm}. Usually, in the $p\rightarrow \infty$ limit,
the one-step RSB yields the correct solution, revealing new phases
which are not present in the corresponding RS approach.  

It should be mentioned that, recently, there has been an increasing interest
in the study of $p$-spin-interaction SG models, motivated either
by attaining a better understanding of such models, or by the possible
applications in other fields of science. The study of infinite-range
$p$-spin-interaction SG models, by means of the replica method, has 
produced many interesting new features as a consequence of the inclusion
of magnetic fields \cite{oliv,gil1}, ferro- \cite{dorl,gil2,gil3}
and antiferromagnetic \cite{jai1,jai2} interactions, as well as a
competition between quadrupolar and SG orderings \cite{jm}. Recent
dynamical studies also have been carried \cite{lopa,sals}, leading,
in particular, to an analysis of the barriers separating metastable
states \cite{lopa}. Besides that, such models have been investigated
lately through rigorous approaches \cite{tala} and in a quantum version
\cite{biro}; new applications were explored, e.g., connections to the
protein folding problem \cite{pand}, to error correcting codes \cite{mert},
and to many biological systems \cite{dros}.

The Ashkin-Teller model \cite{ast} is one of most studied systems in
statistical mechanics, due mainly to the richness of critical phenomena
revealed by its phase diagrams both in two and three dimensions \cite{gbdk}.
This wealth of results steams from the competition between the two- and
four-spin interactions present in the model. Herein we consider an
infinite-range Ashkin-Teller-like SG with $p$-spins interactions in
order to investigate the effects of an analogous competition. We present the
full solution in the $p\rightarrow \infty$ limit. The resulting phase diagram
shows three distinct phases, similarly to what has been found in other
Ashkin-Teller SG models with $p=2$ \cite{cgr,mc,ns}. However, the
nature of the present transition lines are changed, and we get a triple
point common to the three phases, instead of a multicritical one as found
in the previous $p=2$ works \cite{cgr,mc,ns}.
In particular, we find that in the limit of two independent Ising-like models
(or 4-state clock model) the equilibrium transition occurs at a multiphase
point. We also show that, for $p \to \infty$, some particular cases of
the present model are equivalent to
random-energy models with uncorrelated energy levels.

This paper is organized as follows. In the next section we introduce the model
and determine the free-energy density functional, obtained through the replica
method. In section 3 we determine the phase diagram within the RS
solution and consider the stability of such a solution against Gaussian
fluctuations in replica space. In section 4 we apply the
first stage of Parisi's {\em Ansatz} to determine the phase diagram. Our main
conclusions are drawn in section 4. Equivalences with random-energy models,
in the limit $p \to \infty$, are shown in an appendix.

\section{The Model}
In the present work we will consider a $p$-spin interaction Ashkin-Teller-like
SG model, defined by the Hamiltonian

\begin{equation}\label{eq1}
H = - \sum_{1 \leq i_{1} < ... < i_{p} \leq N}
       {\left\{J^{(1)}_{i_{1} ... i_{p}} \sigma_{i_{1}} ... \sigma_{i_{p}}+
          J^{(2)}_{i_{1} ... i_{p}} \tau_{i_{1}} ... \tau_{i_{p}}+
          J^{(4)}_{i_{1} ... i_{p}} \sigma_{i_{1}}\tau_{i_{1}} ...
          \sigma_{i_{p}}\tau_{i_{p}}\right\}},
\end{equation}

\noindent
where $\sigma$ and $\tau$ ($=\pm 1$) are Ising spin variables. All
interactions are infinite-range-like, and similarly
to what has been done in previous
works \cite{cgr,mc}, herein we will be restricted to the case of
independent couplings $J^{(\alpha)}_{i_{1} ... i_{p}}$
($\alpha = 1,2,4$), each one following its own Gaussian probability
distribution,

\begin{equation}\label{eq2}
P\left(J^{(\alpha)}_{i_{1} ... i_{p}}\right) =
     \sqrt{\frac{N^{p-1}}{\pi p{\,}!  J^{2}_{\alpha}}}
       \exp{\left\{-\frac{N^{p-1}{J^{(\alpha)}_{i_{1} ... i_{p}}}^{2}}
                            {p{\,}!  J^{2}_{\alpha}}\right\}} .
\end{equation}

\noindent
The free-energy density is given by

\begin{equation}\label{fed}
\beta f = - \lim_{N \to \infty}{\frac{1}{N} \langle \ln{Z} \rangle}
\end{equation}

\noindent
where $\beta = (k_{B}T)^{-1}$, and $\langle {\quad} \rangle$ represents an
average over the disorder. In the following we will use the
replica method to calculate

\begin{equation}\label{eq6}
\langle \ln{Z} \rangle = \lim_{n \to 0}{\frac{\langle Z^{n} \rangle -
1}{n}} .
\end{equation}

\noindent
Performing the averages over the random couplings, one gets

\begin{equation}
\begin{split}
\langle Z^{n} \rangle =&{\rm Tr}~{\rm exp} \Biggl[ \frac{p{\,}!(\beta J_1)^2}
{4N^{p-1}}\sum_{1 \leq i_{1} < \cdots < i_{p} \leq N}\left( \sum_{a=1}^{n}
\sigma_{i_1}^{a} \sigma_{i_2}^{a} \cdots \sigma_{i_p}^{a}\right)^2  \\
& + \frac{p{\,}!(\beta J_2)^2}
{4N^{p-1}}\sum_{1 \leq i_{1} < \cdots < i_{p} \leq N}\left( \sum_{a=1}^{n}
\tau_{i_1}^{a} \tau_{i_2}^{a} \cdots \tau_{i_p}^{a}\right)^2 \\
& + \frac{p{\,}!(\beta J_4)^2}
{4N^{p-1}}\sum_{1 \leq i_{1} < \cdots < i_{p} \leq N}\left( \sum_{a=1}^{n}
\sigma_{i_1}^{a}\tau_{i_1}^{a} \sigma_{i_2}^{a}\tau_{i_2}^{a} \cdots
\sigma_{i_p}^{a}\tau_{i_p}^{a} \right)^2
\Biggr] ,  \\
\end{split}
\end{equation}
where $a = 1, \ldots , n$ represents the replica index. As usual, the sums
over $p$ sites may be reduced to sums over a single site, e.g.,

\begin{equation}
\sum_{1 \leq i_{1} < \cdots < i_{p} \leq N}\left( \sum_{a=1}^{n}
\sigma_{i_1}^{a} \sigma_{i_2}^{a} \cdots \sigma_{i_p}^{a}\right)^2 =
\sum_{a,b=1}^{n} \sum_{1 \leq i_{1} < \cdots < i_{p} \leq N}
\sigma_{i_1}^{a} \sigma_{i_1}^{b} \sigma_{i_2}^{a} \sigma_{i_2}^{b}
\cdots \sigma_{i_p}^{a} \sigma_{i_p}^{b}
\nonumber
\end{equation}

\vspace{-1cm}

\begin{equation}
={1 \over p{\,}!} \sum_{a,b=1}^{n} \left( \sum_{i}
\sigma_{i}^{a} \sigma_{i}^{b} \right)^{p} + O(N^{p-1})
\nonumber
\end{equation}

\vspace{-1cm}

\begin{equation}
={1 \over p{\,}!} N^{p}n +
{2 \over p{\,}!} N^{p} \sum_{(ab)} \left( {1 \over N} \sum_{i}
\sigma_{i}^{a} \sigma_{i}^{b} \right)^{p} + O(N^{p-1}),
\nonumber
\end{equation}
where $\sum_{(ab)}$ denotes a sum over distinct pairs of replicas.
One may now introduce, for each distinct pair of replicas $(ab)$,
the order parameters,

\begin{align}
q_{1,ab}&=\frac{1}{N}\sum_{i}{\sigma_{i}^{a}\sigma_{i}^{b}}
= \langle \sigma^{a} \sigma^{b} \rangle ,  \tag{6.a} \label{q1ab}\\
q_{2,ab}&=\frac{1}{N}\sum_{i}{\tau_{i}^{a}\tau_{i}^{b}}
= \langle \tau^{a} \tau^{b} \rangle , \tag{6.b} \label{q2ab} \\
r_{ab}&=\frac{1}{N}\sum_{i}{\sigma_{i}^{a}\tau_{i}^{a}\sigma_{i}^{b}
\tau_{i}^{b}}
= \langle \sigma^{a} \tau^{a} \sigma^{b} \tau^{b} \rangle , \tag{6.c}
\label{rab}
\end{align}
\addtocounter{equation}{1}

as well as their respective auxiliary fields (Lagrange's multipliers)
$\gamma_{1,ab}$, $\gamma_{2,ab}$, and $\xi_{ab}$ through standard identities,
e.g., for the pair
$(\mathbf{q_1},\boldsymbol{\gamma_1})$,

\begin{equation}
{\left(\frac{N}{2 \pi \imath}\right)}^{n^{2}}
  \int_{-\infty}^{\infty}{\int_{-\imath \infty}^{\imath \infty}
  {\prod_{(ab)} d\gamma_{1,ab} dq_{1,ab}
e^{L(\{\gamma_{1,ab}\},\{q_{1,ab}\})}}} = 1
\end{equation}
where,
\begin{equation}
L = -\sum_{(ab)}{\gamma_{1,ab}\left(Nq_{1,ab}-\sum_{i}{\sigma_{i}^{a}
            \sigma_{i}^{b}}\right)} .
\end{equation}
Applying a similar procedure for the pairs of matrices $(\mathbf{q_2},
\boldsymbol{\gamma_2})$ and $(\mathbf{r},\boldsymbol{\xi})$, one may write

\begin{eqnarray} \label{lnz}
\langle Z^{n} \rangle & = & {\left(\frac{N}{2 \pi \imath}\right)}^{3n^{2}}
           \int_{-\infty}^{\infty}{\int_{-\imath \infty}^{\imath \infty}
           {\prod_{(ab)} d\gamma_{1,ab} dq_{1,ab} d\gamma_{2,ab}
           dq_{2,ab}d\xi_{ab} dr_{ab}}} \nonumber \\
     &   & \times \exp{\left[-N g_{n}(\{\gamma_{1,ab}\},
\{q_{1,ab}\},\{\gamma_{2,ab}\},\{q_{2,ab}\},\{\xi_{ab}\},\{r_{ab}\})\right]}
\end{eqnarray}
where

\begin{eqnarray} \label{gnf}
g_{n}&=&\sum_{(a,b)}{\left(\gamma_{1,ab}q_{1,ab}+\gamma_{2,ab}q_{2,ab}
     +\xi_{ab}r_{ab}\right)}
     \nonumber \\ & & -\frac{\beta^{2}}{2}\sum_{(a,b)}
{\left(J_{1}^{2}q_{1,ab}^{p}+J_{2}^{2}q_{2,ab}^{p}+J_{4}^{2}r_{ab}^{p}\right)}
    -\frac{\beta^{2}n}{4}\left(J_{1}^{2}+J_{2}^{2}+J_{4}^{2}\right)  \nonumber
\\ & &
 -\ln{{\rm Tr}\exp{\left\{\sum_{(ab)}{\left[\gamma_{1,ab}\sigma_{i}^{a}
    \sigma_{i}^{b}
    +\gamma_{2,ab}\tau_{i}^{a}\tau_{i}^{b}
    +\xi_{ab}\sigma_{i}^{a}\tau_{i}^{a}\sigma_{i}^{b}\tau_{i}^{b}\right]}
    \right\}}} .
\end{eqnarray}

Substituting eq. (\ref{lnz}) into eq. (\ref{eq6}), we obtain the free-energy
density functional

\begin{equation} \label{fef}
\beta f  = \lim_{n \to 0}{\frac{1}{n}}\tilde{g}_{n}
 (\{\gamma_{1,ab}\},\{q_{1,ab}\},\{\gamma_{2,ab}\},\{q_{2,ab}\},\{\xi_{ab}\},\{
r_{ab}\})
\end{equation}

\noindent
where $\tilde{g}_{n}$ stands for the global minimum of $g_n$, taken with
respect to the variational parameters.

One must now look for stable solutions
of the saddle-point equations. The simplest {\em Ansatz} consists in
considering all
replicas in an equal footing, and thus assume the RS
solution. In the next section we will analyse this solution, and it will be
shown that the only stable RS solution is the paramagnetic one.

\section{Replica-Symmetric Solution}

The RS solution is obtained by considering

\begin{equation}
q_{1,ab} = q_{1} , \quad q_{2,ab} = q_{2} , \quad {\rm and}\quad r_{ab} = r,
\quad \forall
\, (ab) , \tag{12.a} \label{eq:rsop1}
\end{equation}
as well as

\begin{equation}
\gamma_{1,ab} = \gamma_{1} , \quad \gamma_{2,ab} = \gamma_{2} , \quad {\rm and}
\quad \xi_{ab} = \xi, \quad \forall \, (ab). \tag{12.b} \label{eq:rsop2}
\end{equation}
\addtocounter{equation}{1}

Performing standard simplifications through Gaussian identities, we get
the free-energy density functional as

\begin{eqnarray}\label{fedrs}
f &=& \frac{\beta}{4}\left[J_{1}^{2}(q_{1}^{p}-1)+J_{2}^{2}(q_{2}^{p}-1)+
  J_{4}^{2}(r^{p}-1)\right] \\
  & & -\frac{1}{2\beta}\left[\gamma_{1} (q_{1}-1)+\gamma_{2} (q_{2}-1)+\xi
(r-1)\right]\nonumber
     -\frac{1}{\beta}{\dla \ln{\Xi (x,y,z)} \dra _{x,y,z}}~,
\end{eqnarray}

\noindent
where

\begin{eqnarray}\label{eq8}
\Xi (x,y,z)  =&4[
\cosh({\sqrt{\gamma_{1}}x})\cosh({\sqrt{\gamma_{2}}y})\cosh({\sqrt{\xi}z})
      \nonumber \\
      & +
\sinh({\sqrt{\gamma_{1}}x})\sinh({\sqrt{\gamma_{2}}y})\sinh({\sqrt{\xi}z})]
      .
\end{eqnarray}

\noindent
In Eq. (\ref{fedrs}) the double brackets
$\dla {\,} \dra _{x,y,z}$ stand for Gaussian averages with respect to the
set of variables ($x,y,z$), e.g.,
for an arbitrary function $\varphi(x,y,z)$, one has

$$\label{eq11}
{\dla \varphi(x,y,z) \dra}_{x,y,z} = \int_{-\infty}^{\infty}
          \int_{-\infty}^{\infty} \int_{-\infty}^{\infty}
                  {\frac{dx dy dz}{{(2 \pi)}^{\frac{3}{2}}}
                  {\rm e}^{\frac{1}{2}(x^{2}+y^{2}+z^{2})} \varphi(x,y,z) } .
$$

The parameters $q_{1}$, $q_{2}$, $r$, $\gamma_{1}$, $\gamma_{2}$ and $\xi$
in Eq. (\ref{fedrs}) may be determined from the saddle-point conditions.
Thus, the equations relating the auxiliary fields
$(\gamma_{1},\gamma_{2},\xi)$ to the order parameters $(q_{1},q_{2},r)$
become
\begin{equation}\label{eqaux}
\gamma_{1}  =  \frac{1}{2}p{(\beta J_{1})}^{2}q_{1}^{p-1} , \quad
\gamma_{2}  =  \frac{1}{2}p{(\beta J_{2})}^{2}q_{2}^{p-1} ,  \quad
\xi  =  \frac{1}{2}p{(\beta J_{4})}^{2}r^{p-1} ,
\end{equation}

\noindent
whereas the order parameters are self-consistently given by
\begin{equation}\label{eqop}
q_{1} = {\dla \varphi_{1}^{2}(x,y,z) \dra}_{x,y,z}~, \quad
q_{2} = {\dla \varphi_{2}^{2}(x,y,z) \dra}_{x,y,z}~, \quad
r = {\dla \varphi_{3}^{2}(x,y,z) \dra}_{x,y,z}~,
\end{equation}

\noindent
with

\begin{gather}
\varphi_{1}(x,y,z) = \frac{1}{D}\bigl[\tanh({\sqrt{\gamma_{1}}x})
+ \tanh({\sqrt{\gamma_{2}}y})\tanh({\sqrt{\xi}z})\bigr] , \tag{17.a} \\
\varphi_{2}(x,y,z) = \frac{1}{D}\bigl[\tanh({\sqrt{\gamma_{2}}y})
+ \tanh({\sqrt{\gamma_{1}}x})\tanh({\sqrt{\xi}z})\bigr] , \tag{17.b}\\
\varphi_{3}(x,y,z) = \frac{1}{D}\bigl[\tanh({\sqrt{\xi}z})
+ \tanh({\sqrt{\gamma_{1}}x})\tanh({\sqrt{\gamma_{2}}y})\bigr] , \tag{17.c}
\end{gather}
\addtocounter{equation}{1}

\noindent
and $D = 1 + \tanh({\sqrt{\gamma_{1}}x}) \tanh({\sqrt{\gamma_{2}}y})
\tanh({\sqrt{\xi}z})$.

It would be interesting to investigate the full phase diagram
obtained from the above equations. However, similar to what has been done
in previous studies of $p=2$ Ashkin-Teller
models, the most interesting case is the
isotropic one, for which $J_{2} = J_{1}$. In such a case, besides the usual
spin reversal symmetries, there is a new symmetry in which the $\sigma$
and $\tau$ spins variables may be interchanged.

For $p > 1$ there is always a disordered, paramagnetic, phase where $q_{1}
= q_{2} = r = 0$ and $\gamma_{1} = \gamma_{2} = \xi = 0$, with the free energy
given by

\begin{equation}\label{FP}
f_{\mathrm P} = -\frac{1}{4T}(2 J_2^2 + J_4^2) - 2T \ln {2} ,
\end{equation}
where $T$ is the temperature (herein we work in units such that
$k_{\mathrm B} = 1$).
From the above expression we obtain the entropy density

\begin{equation}\label{sP}
s_{\mathrm P} = -\frac{1}{4T^2}(2J_2^2 + J_4^2) + 2\ln {2}.
\end{equation}

\begin{figure}
\begin{center}
\includegraphics[width=8cm]{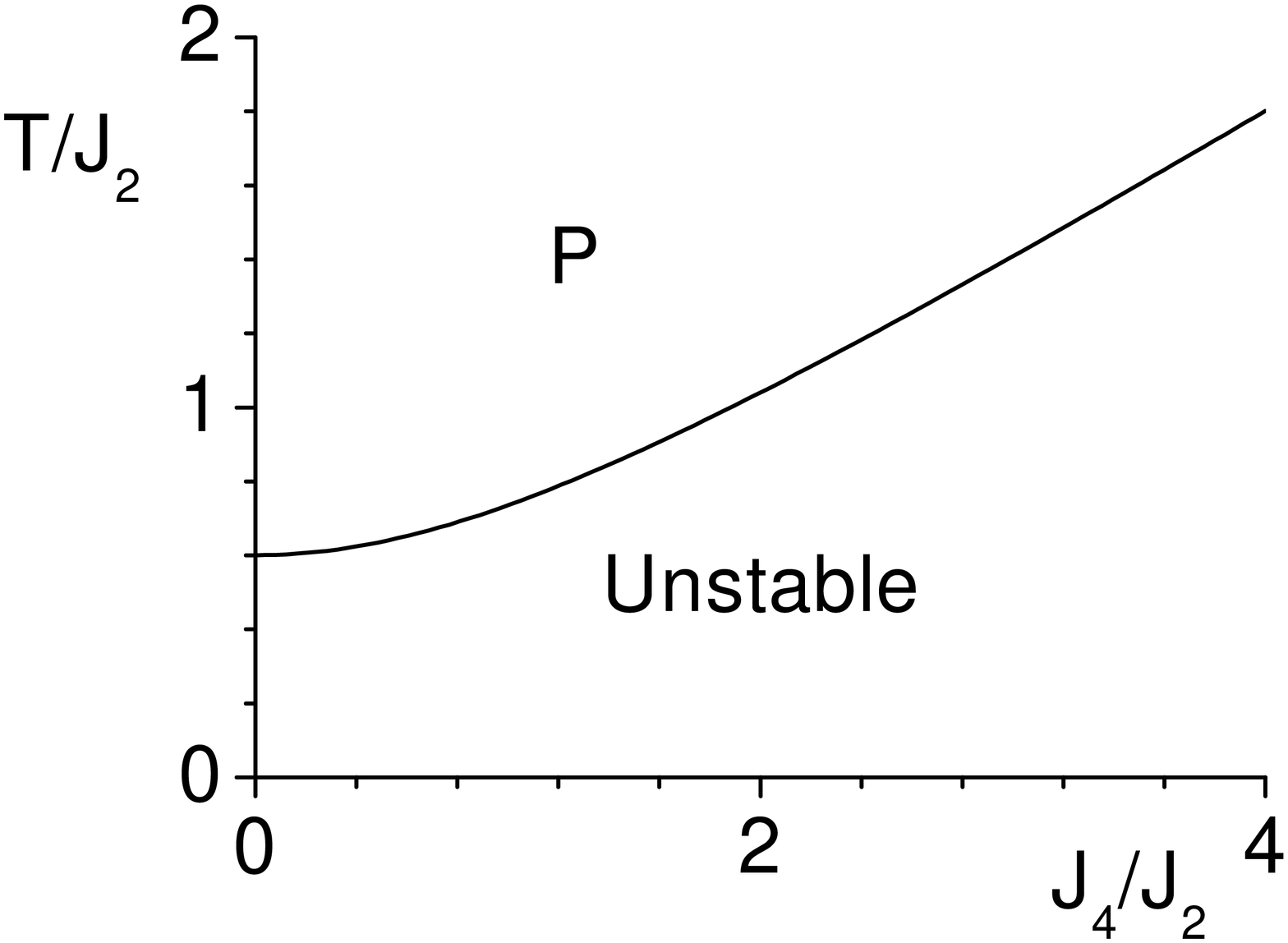}
\end{center}
\caption{\footnotesize{Phase diagram of the $p$-spin interaction
Ashkin-Teller SG, within the RS solution, for arbitrary values of
$p$. At low temperatures, the paramagnetic solution (P) becomes unstable,
and there are no stable non-trivial solutions.}}
\label{fig1}
\end{figure}

For low temperatures this entropy becomes negative and
the system gets frozen. The paramagnetic phase can exist only above the curve
shown in Fig. \ref{fig1}, which is given by

\begin{equation}
\frac{T}{J_2} = \frac{1}{2\sqrt{\ln 2}}\left(1 + \frac{J_4^2}{2J_2^2}
\right)^{1/2}.
\end{equation}

There are many other non-trivial RS solutions.
The stability analysis of those solutions against replica fluctuations can
be performed along the lines pioneered by de Almeida and Thouless
\cite{alt} for the $p=2$ infinite-range Ising SG model. In general, the
eigenvalues are roots of cubic equations, and become too lengthy to quote
their expressions, except in a few particular cases. For instance, to the
$q_1 = q_2 = 0$, $\gamma_1 = \gamma_2 = 0$, $r \neq 0$ solution there
corresponds, in the limit $n \rightarrow 0$, the following {\em longitudinal}

\begin{equation}\label{eq13}
\lambda_{L}=\frac{1}{2}{\left(\beta J_{4}\right)}^{2}p(p-1)r^{p-2}
                  \left[1-\frac{1}{2}{\left(\beta
J_{4}\right)}^{2}p(p-1)r^{p-2}
                  \left(1-4r+3t\right)\right] ,
\end{equation}
and {\em transversal}
\begin{equation}
\lambda_{T}=\frac{1}{2}{\left(\beta J_{4}\right)}^{2}p(p-1)r^{p-2}
                  \left[1-\frac{1}{2}{\left(\beta
J_{4}\right)}^{2}p(p-1)r^{p-2}
                  \left(1-2r+t\right)\right]
\end{equation}

eigenvalues, where
\begin{equation}
t = \dla \varphi_1^4(x,y,z) \dra _{x,y,z}~.
\end{equation}
When $r =0$, one gets the paramagnetic solution
and, from the vanishing of the above eigenvalues, such a solution
is marginally stable for $p>2$. On the other hand, in the $p \to \infty$
the $r=1$ solution presents both eigenvalues negative, being completely
unstable. The same happens for any other non-trivial RS solution. Thus, we
must look for RSB solutions; this will be done in the next section.

\section{Replica Symmetry Breaking Solution}

\noindent
Since there is no stable SG solution in the $p \to \infty$ limit, within
the RS assumption, we
need to look for RSB solutions. Fortunately, it is enough to consider just
the first step in Parisi's RSB procedure in order to get the correct
solution in such a limit, as verified in other $p$-spin-interaction models
studied before \cite{gm,mot,jm,gar}. In
the present case, since we deal with three distincts pairs of
conjugated matrices,
namely $(\mathbf{q_1},\boldsymbol{\gamma_1})$,
$(\mathbf{q_2},\boldsymbol{\gamma_2})$, and
$(\mathbf{r},\boldsymbol{\xi})$, one should apply the RSB scheme for all
six matrices.
In principle, one could
apply the RSB scheme for each matrix in an independent way, i.e.,
each matrix should have its own block sizes.
However, it is easy to convince oneselves that
this is not a physically acceptable procedure, due to the symmetries involving
the interchanging between the spins in the Hamiltonian. Therefore, herein
we will apply the same RSB for each matrix, i.e., we will divide all six
$n \times n$ matrices into $n/m$ groups of size $m$.
Following Parisi \cite{par}, we denote the elements
of the off-diagonal blocks by $q_{1,0}, q_{2,0}, r_0, \gamma_{1,0},
\gamma_{2,0}$ and ${\xi_0}$, and those of the diagonal blocks
by $q_{1,1}, q_{2,1}, r_1, \gamma_{1,1}, \gamma_{2,1}$ and ${\xi_1}$.
Thus, we obtain the free-energy density functional

\begin{eqnarray}\label{eq14}
f &=& -\frac{\beta}{4}(J_{1}^{2}q_{1,1}^{p}+J_{2}^{2}q_{2,1}^{p}+
J_{4}^{2}r_{1}^{p})(m-1)
  +\frac{\beta m}{4}(J_{1}^{2}q_{1,0}^{p}+J_{2}^{2}q_{2,0}^{p}+
J_{4}^{2}r_{0}^{p})
\nonumber \\
&&+\frac{1}{2\beta}(\gamma_{1,1}q_{1,1}+\gamma_{2,1}q_{2,1}
+\xi_{1}r_{1})(m-1)
  -\frac{m}{2\beta}(\gamma_{1,0}q_{1,0}+\gamma_{2,0}q_{2,0}
  +\xi_{0}r_{0})
\nonumber
\\
&&-\frac{\beta}{4}(J_{1}^{2}+J_{2}^{2}+ J_{4}^{2})
  +\frac{1}{2\beta}(\gamma_{1,1}+\gamma_{2,1}+ \xi_{1}) \nonumber \\
&&-\frac{1}{\beta
m}\dla \ln Z(x_0,x_1,x_2) \dra _{x_{0},x_{1},x_{2}}~,
\end{eqnarray}

\noindent where
\begin{eqnarray}\label{eq15}
Z(x_0,x_1,x_2)=  \dla A^{m}(x_0,x_1,x_2,y_0,y_1,y_2) \dra _{y_{0},y_{1},y_{2}}~,
\end{eqnarray}

\begin{eqnarray}\label{eq16}
A&=&4\cosh(u_0)\cosh(u_1)\cosh(u_2)+ 4\sinh(u_0)\sinh(u_1)\sinh(u_2) ,
\end{eqnarray}
with $u_0$, $u_1$ and $u_2$ defined by
\begin{eqnarray}\label{eq17}
u_{0}&=&\sqrt{\gamma_{1,0}} \ x_{0}+\sqrt{\gamma_{1,1}-\gamma_{1,0}} \ y_{0} ~ ,
\qquad
u_{1}=\sqrt{\gamma_{2,0}} \ x_{1}+\sqrt{\gamma_{2,1}-\gamma_{2,0}} \ y_{1} ~ ,
\nonumber \\
u_{2}&=&\sqrt{\xi_{0}} \ x_{2}+\sqrt{\xi_{1}-\xi_{0}} \ y_{2} ~ .
\end{eqnarray}

\begin{figure}
\begin{center}
\includegraphics[width=8cm]{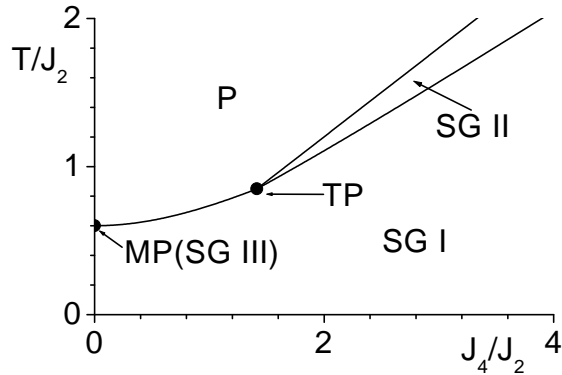}
\end{center}
\caption{\footnotesize{Phase diagram of the $p$-spin interaction
Ashkin-Teller SG, within a one-step RSB procedure, in the limit
$p \rightarrow \infty$. The borders of the paramagnetic phase (P) present
discontinuites in the order parameters, but no latent heat, whereas the
critical frontier SGI-SGII is a genuine first-order phase transition, with
discontinuities in the order parameters and a finite latent heat. The phases
P, SGI, and SGII coexist at a triple point (TP), following the standard Gibbs
phase rule. A distinct SG solution (SGIII) becomes possible at the
multiphase point (MP), where the phases P, SGI, and SGIII coexist.}}
\label{fig2}
\end{figure}

\noindent
By extremizing the above free-energy density one gets the equations of state
within the one-step RSB procedure (see Appendix A); it is easy to see that
such equations present several non-trivial solutions in the
limit $p \rightarrow \infty$. A careful
analysis of the free energy shows that only four of those solutions
are physically acceptable. For low temperatures there is always a SGI
phase (see Fig. \ref{fig2}), in which $q_{1,1} = q_{2,1} = r_1 = 1$,
$q_{1,0} = q_{2,0} = r_0 = 0$, and

\begin{equation}
m = \frac{2\sqrt{2 \ln 2}}{\sqrt{2+(J_4/J_2)^2}}\frac{T}{J_2} .
\end{equation}
The corresponding free-energy density is given by

\begin{equation} \label{FI}
f_{I} = -\sqrt{(4J_2^2 + 2J_4^2)\ln 2} ~,
\end{equation}
which yields a vanishing entropy. The transition between the SGI
phase and the paramagnetic one occurs along a line given by

\begin{equation} \label{FOT1}
\frac{T}{J_2} = \frac{1}{2\sqrt{2 \ln 2}}\sqrt{2 + (J_4/J_2)^2} ~~ ,
\end{equation}
where both phases present zero entropy, i.e., there is no latent heat.
This line holds as long as $J_4/J_2 \le \sqrt{2}$. Beyond that point,
we find two other first-order transition lines confining a
SGII phase, as shown in Fig. \ref{fig2}, throughout which one has
$r_1 = 1$, $m = 2\sqrt{\ln 2} \ T/J_4$,
and all other order parameters zero. This phase presents a free-energy
density

\begin{equation} \label{FII}
f_{II} = -\sqrt{\ln 2} \ J_4 - \frac{J_2^2}{2T} - T \ln 2 ~,
\end{equation}

\noindent which leads to the entropy density

\begin{equation} \label{SII}
s_{II}(T) = - \frac{J_2^2}{2T^2} + \ln 2 ~.
\end{equation}
Due to discontinuities in the order parameters, the transition
separating the paramagnetic and SGII phases is first-order;
this critical frontier may be determined by demanding the free-energy
densities given by Eqs. (\ref{FP}) and (\ref{FII}) to be equal, leading
to the straight line,

\begin{equation} \label{FOT2}
T = \frac{J_4}{2\sqrt{\ln 2}} ~.
\end{equation}
Along this critical frontier there is also no latent heat.
For lower temperatures another transition occurs, this time from
the SGII to the SGI phase. From their free-energy densities, given
respectively by Eqs. (\ref{FII}) and (\ref{FI}), we find the corresponding
critical frontier,

\begin{equation} \label{FOT3}
\frac{T}{J_2} = \frac{1}{2\sqrt{\ln 2}} \left[(2+\sqrt{2})\sqrt{1 +
\frac{J_4^2}{2J_2^2}} - (1 + \sqrt{2})\frac{J_4}{J_2} \right] .
\end{equation}

It is easy to verify that along the SGI-SGII critical frontier
there is a finite latent
heat. Therefore, in this case one has a genuine first-order transition
line. The three transition lines given by eqns. (\ref{FOT1}),
(\ref{FOT2}) and (\ref{FOT3}) merge together at a triple point (TP), located
at $J_4/J_2=\sqrt{2}$ and $J_2/T = \sqrt{2\ln{2}}$, as shown in
Fig. \ref{fig2}. Although two of
these transition lines do not represent conventional first-order
transitions in the thermodynamic sense, since we do not observe any
discontinuity in the first derivatives of their corresponding
free-energy densities, i.e., no latent heat, the picture around the
triple point follows the standard Gibbs phase rule \cite{callen}.

Besides the solutions discussed above, there are also two equivalent
solutions (which we call SGIII):
$q_{1,1} = 1$, $m = 2\sqrt{\ln 2} \ T/J_2 $, with all
remaining order parameters zero, and its symmetric one, i.e.,
$q_{2,1} = 1$, $m = 2\sqrt{\ln 2} \ T/J_2$, with all remaining order
parameters equal to zero. However, the SGIII phase is realized only at
the multiphase point (MP), $[J_4 = 0, T/J_2 = 1/(2 \sqrt{\ln 2})]$, where it
coexists with the paramagnetic and SGI phases.

\section{Conclusion}

\noindent
We have studied a version of the
Ashkin-Teller SG model, with $p$-spin interations, by means of the
replica approach. The RS solution leads to a simple phase diagram, for
arbitrary values of $p$, with a
paramagnetic phase that is stable at high temperatures, becoming unstable
at low temperatures; within such a solution, there are no stable non-trivial
solutions at low temperatures. By applying a one-step RSB procedure, we have
found a rich phase diagram in the
$p \to \infty$ limit, with four distinct phases, namely, a paramagnetic one
at high temperatures, and three SG phases (SGI, SGII, and SGIII)
at lower temperatures.
The borders of the paramagnetic phase present discontinuities in the order
parameters, but no latent heat, whereas the critical frontier separating
phases SGI and SGII is a genuine first-order phase transition, exhibiting
both discontinuities in the order parameters, as well as a finite latent
heat. These critical frontiers all meet at a triple point according to the
standart Gibbs phase rule \cite{callen}, similarly to what happens in a
previously investigated model \cite{jm}. The SGIII solution is stable only
at a multiphase
point, where it coexists with the paramagnetic and SGI phases.
Also, in the $p \to \infty$ limit, it is possible to show the equivalence
of the model considered herein with a random energy model, as defined by
Derrida \cite{der80,der81}, which can be solved by other methods; this
equivalence is shown in Appendix B. A detailed analysis of the
corresponding random-energy model will be published elsewhere.


\section*{Acknowledgments}

\noindent We would like to thank CNPq (Brazilian Government) for
financial support.


\appendix
\section{Equations of State for the Replica-Symmetry-Breaking Solution}

\noindent
In this appendix we derive the equations of state
of the one-step RSB solution discussed in section 4.
We shall consider the same matrix block sizes,
i.e., the same value of $m$ for all matrices; denoting the elements
of the off-diagonal blocks by $q_{1,0}, q_{2,0}, r_0, \gamma_{1,0},
\gamma_{2,0}$ and ${\xi_0}$, whereas those of the diagonal blocks
by $q_{1,1}, q_{2,1}, r_1, \gamma_{1,1}, \gamma_{2,1}$ and
${\xi_1}$, one gets the free-energy density defined in
Eqs. (\ref{eq14}) -- (\ref{eq17}). The extremization of such a free-energy density
with respect to this set of parameters leads to the following equations
of state,

\begin{eqnarray}
\gamma_{1,0}&=&\frac{1}{2}p{\left(\beta J_{1}\right)}^{2}q_{1,0}^{p-1}
\\
\gamma_{1,1}&=&\frac{1}{2}p{\left(\beta J_{1}\right)}^{2}q_{1,1}^{p-1}
\\
\gamma_{2,0}&=&\frac{1}{2}p{\left(\beta J_{2}\right)}^{2}q_{2,0}^{p-1}
\\
\gamma_{2,1}&=&\frac{1}{2}p{\left(\beta J_{2}\right)}^{2}q_{2,1}^{p-1}
\\
\xi_{0}&=&\frac{1}{2}p{\left(\beta J_{4}\right)}^{2}r_{0}^{p-1}
\\
\xi_{1}&=&\frac{1}{2}p{\left(\beta J_{4}\right)}^{2}r_{1}^{p-1}
\end{eqnarray}

\begin{eqnarray}
q_{1,0}&=&\bigdla \left[ { \dla A^{m-1}B \dra _{y_{0},y_{1},y_{2}}
\over \dla A^{m} \dra _{y_{0},y_{1},y_{2}} } \right]^{2}
\bigdra _{x_{0},x_{1},x_{2}}  \\
q_{1,1}&=&\bigdla { \dla A^{m-2}B^{2} \dra _{y_{0},y_{1},y_{2}}
\over \dla A^{m} \dra _{y_{0},y_{1},y_{2}} }
\bigdra _{x_{0},x_{1},x_{2}}  \\
q_{2,0}&=&\bigdla \left[ { \dla A^{m-1}C \dra _{y_{0},y_{1},y_{2}}
\over \dla A^{m} \dra _{y_{0},y_{1},y_{2}} } \right]^{2}
\bigdra _{x_{0},x_{1},x_{2}}  \\
q_{2,1}&=&\bigdla { \dla A^{m-2}C^{2} \dra _{y_{0},y_{1},y_{2}}
\over \dla A^{m} \dra _{y_{0},y_{1},y_{2}} }
\bigdra _{x_{0},x_{1},x_{2}}  \\
r_{0}&=&\bigdla \left[ { \dla A^{m-1}D \dra _{y_{0},y_{1},y_{2}}
\over \dla A^{m} \dra _{y_{0},y_{1},y_{2}} } \right]^{2}
\bigdra _{x_{0},x_{1},x_{2}}  \\
r_{1}&=&\bigdla { \dla A^{m-2}D^{2} \dra _{y_{0},y_{1},y_{2}}
\over \dla A^{m} \dra _{y_{0},y_{1},y_{2}} }
\bigdra _{x_{0},x_{1},x_{2}}~,
\end{eqnarray}

\noindent
where we have used the notation introduced in section 3, i.e.,
the double brackets $\dla {\,} \dra _{x,y,z}$ stand for Gaussian averages
with respect to the set of variables ($x,y,z$).
The extremization with respect to the parameter associated with the block
sizes, $m$, leads to

\begin{eqnarray}
0&=&-\frac{\beta}{4}\left[J_{1}^{2}\left(q_{1,1}^{p}-q_{1,0}^{p}\right)
  +J_{2}^{2}\left(q_{2,1}^{p}-q_{2,0}^{p}\right)
  +J_{4}^{2}\left(r_{1}^{p}-r_{0}^{p}\right)\right] \nonumber \\
&&+\frac{1}{2\beta}\left(\gamma_{1,1}q_{1,1}-\gamma_{1,0}q_{1,0}
  +\gamma_{2,1}q_{2,1}-\gamma_{2,0}q_{2,0}+\xi_{1}r_{1}-\xi_{0}r_{0}\right)
\nonumber
\\
&&+\frac{1}{\beta m^{2}}
\bigdla \ln \dla A^{m} \dra _{y_{0},y_{1},y_{2}}
\bigdra _{x_{0},x_{1},x_{2}}  \nonumber \\
&&-\frac{1}{\beta m}
\bigdla \left[ { \dla A^{m}\ln A \dra _{y_{0},y_{1},y_{2}}
\over \dla A^{m} \dra _{y_{0},y_{1},y_{2}} } \right]^{2}
\bigdra _{x_{0},x_{1},x_{2}}
\end{eqnarray}

\noindent
In the equations above one has,

\begin{eqnarray}
A&=&4\cosh(u_{0})\cosh(u_{1})\cosh(u_{2})+4\sinh(u_{0})\sinh(u_{1})\sinh(u_{2})
\\
B&=&4\sinh(u_{0})\cosh(u_{1})\cosh(u_{2})+4\cosh(u_{0})\sinh(u_{1})\sinh(u_{2})
\\
C&=&4\cosh(u_{0})\sinh(u_{1})\cosh(u_{2})+4\sinh(u_{0})\cosh(u_{1})\sinh(u_{2})
\\
D&=&4\cosh(u_{0})\cosh(u_{1})\sinh(u_{2})+4\sinh(u_{0})\sinh(u_{1})\cosh(u_{2})
\end{eqnarray}

\noindent
with

\begin{eqnarray}
u_{0}&=&\sqrt{\gamma_{1,0}}\ x_{0}+\sqrt{\gamma_{1,1}-\gamma_{1,0}}\ y_{0}
\\
u_{1}&=&\sqrt{\gamma_{2,0}}\ x_{1}+\sqrt{\gamma_{2,1}-\gamma_{2,0}}\ y_{1}
\\
u_{2}&=&\sqrt{\xi_{0}}\ x_{2}+\sqrt{\xi_{1}-\xi_{0}}\ y_{2}~.
\end{eqnarray}


\section{Equivalence with Random Energy Model}

\noindent
In this appendix we consider two particular cases of the model defined in
Eq. (1), in the limit $p \rightarrow \infty$, and find the corresponding
equivalences with random-energy models. Let us first
consider spin configurations for which $\{\sigma_i\}$ and $\{\tau_i\}$ are
completely independent of each other; this is expected to be valid at high
temperatures, and the corresponding random-energy model should yield the
same results as the model of Eq. (1) in the paramagnetic phase.
The probability that such a configuration presents a given energy $E$ is

\begin{equation}
P_1(E) = \langle \delta(E-{\mathcal H}(\{\sigma_i,\tau_i\}) \rangle ,
\end{equation}
where the average is taken over all possible realizations of $J_{i_1
\ldots i_p}$. Using Eqs. (\ref{eq1}) and (\ref{eq2}) we find

\begin{equation} \label{pd1}
P_1(E) = \frac{1}{\sqrt{N\pi J_{eff}^2}}
\exp\left(-\frac{E^2}{NJ_{eff}^2} \right) ,
\end{equation}
where we have introduced the effective variance $J_{eff}^2 = J_1^2 + J_2^2
+ J_4^2 $. The total number of such configurations is $4^N$; therefore,
the average
number of macroscopic states with energies between $E$ and $E+dE$ is given
by

\begin{equation}  \label{n1e}
n_1(E) = 4^N P_1(E) \approx \exp \left[N\left( 2\ln 2 -
\frac{E^2}{NJ_{eff}^2}\right)
\right] .
\end{equation}
In the thermodynamic limit, $N \to \infty$, it is convenient to
introduce the
energy $ u = E/N  $, and entropy $ s(u) = S(E)/N $  densities. From
Eq. (\ref{n1e})
we obtain

\begin{equation}
s(u) = - \frac{u^2}{J_{eff}^2} + 2 \ln 2 .
\end{equation}
Using the thermodynamic definition of temperature $1/T = \partial s /
\partial u $ we may write

\begin{equation}
s(T) = - \frac{J_{eff}^2}{4T^2} + 2 \ln 2 .
\end{equation}
When $J_1 = J_2$ we recover the entropy density of the paramagnetic phase,
given by Eq. (\ref{sP}).

\noindent
Let us now consider another particular case, namely, situations in which
the $\{\sigma_i\}$
and $\{\tau_i\}$ configurations are the same; this is expect to hold
throughout the SGII phase.
In this case, the last term in (\ref{eq1})
does not contribute, and we get the simplified Hamiltonian

\begin{equation}\label{ham2}
{\mathcal H} = - \sum_{1 \leq i_{1} < ... < i_{p} \leq N}
       \left[J^{(1)}_{i_{1} ... i_{p}} +
          J^{(2)}_{i_{1} ... i_{p}} \right]\sigma_{i_{1}} ...
\sigma_{i_{p}}~.
\end{equation}
The number of such configurations is $2^N$. The probability that one of these
configurations presents an energy $E$ is given by

\begin{equation}
P_2(E) = \langle \delta(E-{\mathcal H}(\{\sigma_i\}) \rangle ,
\end{equation}
with ${\mathcal H}$ given by Eq. (\ref{ham2}). From the above expression and
the probability distribution for the couplings, we find

\begin{equation}  \label{pd2}
P_2(E) = \frac{1}{\sqrt{N\pi (J_1^2+J_2^2)}}
\exp\left[-\frac{E^2}{N(J_1^2+J_2^2)}
\right] .
\end{equation}
Therefore, the mean number of such configurations is

\begin{equation}  \label{n2e}
n_2(E) = 2^N P_2(E) \approx \exp \left\{N\left[ \ln 2 -
\frac{E^2}{N(J_1^2+J_2^2)}\right]
\right\} .
\end{equation}
For $J_1 = J_2$, the entropy density as a function of the energy
density is given by

\begin{equation}
s(u) = - \frac{u^2}{2J_2^2} +  \ln 2 .
\end{equation}
Expressing this entropy as a function of the temperature, one gets the same
entropy density of the phase SGII [cf. Eq. (\ref{SII})],
\begin{equation}
s_{II}(T) = - \frac{J_2^2}{2T^2} +      \ln 2 .
\end{equation}
In summary, we have found the equivalence of our model with random-energy
models in two particular cases:
(i) spin configurations $\{\sigma_i\}$ and $\{\tau_i\}$ independent from
each other. This leads to a
random-energy model, appropriated for the paramagnetic phase, with
$4^N$ independent random-energy levels $E_i$
following the distribution given by (\ref{pd1});
(ii) the same spin configurations $\{\sigma_i\}$ and $\{\tau_i\}$. In this
case one gets a random-energy model, appropriated for the SGII phase, with
$2^N$ independent random-energy levels $E_i$ distributed according to
Eq. (\ref{pd2}). To obtain the free-energy density for the phase SGII
we can use the method introduced by Derrida \cite{der81}, which avoids
the use of replicas.






\end{document}